\documentclass[12pt]{article}
\usepackage{mathrsfs,bm,color}
\usepackage{amsmath}
\usepackage{amssymb}
\newtheorem{theorem}{Theorem}[section]
\newtheorem{df}[theorem]{\bf Definition}
\newtheorem{thm}[theorem]{\bf Theorem}
\newtheorem{cor}[theorem]{\bf Corollary}
\newtheorem{lem}[theorem]{\bf Lemma}

\newtheorem{prop}[theorem]{\bf Proposition}
\newtheorem{assumption}[theorem]{\bf Assumption}
\numberwithin{equation}{section}
\newcommand{\proof}{{\noindent \it Proof:\ }}
\newcommand{\supp}{\mathop{\mathrm{supp}}}

\newcommand{\qed}{\hfill\hbox{\rule{6pt}{6pt}}}

\newcommand{\ov}[1]{\overline{#1}}

\newcommand{\lk}{\left(\!}
\newcommand{\rk}{\!\right)}

\newcommand{\BR}{\RR^d}
\newcommand{\RR}{\mathbb R}
\newcommand{\CC}{\mathbb C}
\newcommand{\ir}{{\rm I}_R}
\newcommand{\HHH}{H}
\newcommand{\EEE}{\widehat  H}
\newcommand{\A}{\ms A}
\renewcommand{\AA}{{\ms A_{E}}}
\renewcommand{\j}{{\rm j}}

\renewcommand{\d}{\displaystyle}
\newcommand{\add}{a^{\dagger}}

\newcommand{\LR}{{L^2(\BR)}}

\newcommand{\Ebb}{\mathbb E}
\newcommand{\ix}{\int_{\BR}\!\!\! {\rm d}x\Ebb^x}

\newcommand{\fff}{{\mathscr{F}}}
\newcommand{\ms}{\mathscr} 
\newcommand{\hhh}{{\mathscr{H}}}

\newcommand{\hf}{{\rm H}_{\rm f}}

\newcommand{\gr}{\Phi_M}

\newcommand{\half}{\frac{1}{2}}

\newcommand{\han}{{1/2}}

\newcommand{\non}{\nonumber}
\newcommand{\bi}{\begin{description}}
\newcommand{\ei}{\end{description} }

\def\bbbone{{\mathchoice {\rm 1\mskip-4mu l} {\rm 1\mskip-4mu l}
{\rm 1\mskip-4.5mu l} {\rm 1\mskip-5mu l}}}
\def\one{\bbbone}
\newcommand{\vp}{\hat\varphi}
\newcommand{\tvp}{\tilde \varphi}

\providecommand{\eq}[1]{\begin{equation}\label{#1}}
\providecommand{\en}{\end{equation}}

\providecommand{\kak}[1]{(\ref{#1})}

\title{\sc Spectrum of 
 the semi-relativistic Pauli-Fierz model I}
\author{\small
Takeru Hidaka
\thanks{
Faculty of Mathematics,
Kyushu University, Fukuoka 819-0385, Japan.
This work is supported  by Grant-in-Aid for JSPS Fellows 24$\cdot$5165.
}
 and  Fumio Hiroshima
\thanks{
Faculty of Mathematics,
Kyushu University, Fukuoka 819-0385, Japan. 
This work is supported by  Grant-in-Aid for Science Research (B) 23340032.
}
}
\date{}
\begin{document}
\maketitle
\begin{abstract}
HVZ type theorem for 
semi-relativistic Pauli-Fierz Hamiltonian, 
$$\HHH=\sqrt{(p\otimes \one -A)^2+M^2}+V\otimes \one +\one\otimes \hf,\quad M\geq 0,$$
in quantum electrodynamics 
is studied. 
Here $H$ is a self-adjoint operator in Hilbert space $\LR\otimes \fff\cong \int^\oplus_{\RR^d}\fff {\rm d}x$, and  $A=\int^\oplus_{\RR^d} A(x) {\rm d}x$   a quantized radiation field and $\hf$ the free field Hamiltonian defined by the second quantization of a  dispersion relation 
$\omega:\RR^d\to \RR$. 
It is emphasized that massless case, 
$M=0$,  is  included. 
Let $E=\inf \sigma (\HHH)$ be the bottom of the spectrum of $\HHH$. 
Suppose that the infimum of $\omega$
 is  $m>0$. Then it is shown that 
$\sigma_{\rm ess}(\HHH)=[E+m, \infty)$. 
In particular 
the existence of the ground state of $\HHH$
can be  proven. 
 \end{abstract}

\section{Introduction}
It is of interest to know  the spectrum of the so-called semi-relativistic Pauli-Fierz model 
 (it is 
shorthanded as the SRPF model) 
in quantum electrodynamics. 
The aim of this paper is to  
specify the essential spectral  of the 
SRPF Hamiltonian.
In the mathematically rigorous quantum field theory spectrum of various models have been investigated so far. 
In particular special attentions have  been  payed for investigating 
the bottom of the spectrum, continuous spectrum and resonances etc. 
The SRPF  model is one of interesting models in quantum electrodynamics.  

The Pauli-Fierz model is a model in non-relativistic quantum electrodynamics 
and 
describes a minimal interaction between electrons 
 governed by a Schr\"odinger operator 
 $\d \frac{1}{2M}p^2+V$,  and a quantized radiation field $A(x)$  
 with an ultraviolet cutoff, which is a self-adjoint operator 
 on the Hilbert space 
\eq{hilebrt}
\hhh=\LR\otimes\fff
\en
 and of the form 
\eq{pfmodel}
H_{PF}=\frac{1}{2M}
\lk 
p\otimes\one-A
\rk^2+V\otimes\one +\one\otimes \hf,
\en
where $p=(-i\partial_{x_1},\cdots,-i\partial_{x_d})$  denotes the $d$-dimensional momentum operator 
of an electron, 
$V$ an external potential, 
$\hf$ the free field Hamiltonian on a Boson Fock space $\fff$, and 
$$\d A=\int_{\BR}^\oplus\!\! A(x) {\rm d}x$$ is the constant fiber direct integral of $A(x)$ under the identification 
$\hhh
\cong\int_{\BR}^\oplus\! \fff {\rm d}x$. 
On the other hand the SRPF model describes a minimal interaction 
between 
$A(x) $ and 
 an electron  governed by semi-relativistic Schr\"odinger operator $\sqrt{p^2+M^2}+V$. 
The total Hamiltonian of the SRPF model is then formally given by 
\eq{HHH}
\HHH =\sqrt{\lk 
p\otimes\one-A
 \rk^2+M^2}+V\otimes\one +\one\otimes \hf.
\en
We give the explicit definition of $\HHH$ later. 
The problems we consider in this paper are 
\bi
\item[1.] HVZ type theorem for  $\HHH $,
\item[2.] the existence and uniqueness of the ground state of $\HHH $.
 \ei
We emphasis that  all the results we obtain in this paper include the case of $M=0$, i.e., 
\eq{h2}
\left|
p\otimes\one-A
\right |
+V\otimes\one +\one\otimes \hf.
\en
Here $|T|=\sqrt{T^2}$ for a self-adjoint operator $T$. 
The crucial  point  is the form of $\d \left|
p\otimes\one-A
 \right |$. 
It is worth pointing out that $x\to |x|$ is not smooth.

We  consider HVZ-type theorem for $\HHH $.
The standard HVZ theorem identifies the essential spectrum of  
$N$-body Schr\"odinger operators. 
See e.g. \cite{Hu}.
We extend HVZ theorem to $\HHH $. 
I.e., we specify the essential spectrum of $\HHH $. 
The bottom of the spectrum of $\HHH $,
 $E $,   is called the ground state energy, and eigenvectors associated with $E $ 
   are called  ground states.
We suppose that a dispersion relation has a strictly positive lower bound $m> 0$. 
Then we shall show that 
\eq{hvz}
\sigma_{\rm ess}(\HHH )=[E +m,\infty).
\en
In particular it can be seen that the gap $m$ is independent of 
the cutoff function in $A(x)$ and $M$, 
and furthermore it is shown that 
$\HHH $ has ground states for all $M\geq0$. 
The method to show this is a combination of 
checking the binding condition developed in 
\cite{GLL01} and functional integration established in \cite{hir97}. 

We review several papers  related to our results.
For the  Pauli-Fierz model $H_{PF}$ 
the existence and uniqueness of ground states 
are proven  in e.g., \cite{BFS, GLL01, hir99}.
For the semi-relativistic case, $\HHH $,  
the existence of a ground state  is  shown in \cite{KMS11, KM13} but for $M>0$. 
In the case of $M=0$ as far as we know 
 however there is no results on the existence of ground states. So our result is new. 
 When $V=0$, 
 $\HHH $ is translation invariant and has no ground state. 
 It can be  however decomposed by the total momentum:
 $$\HHH =\int_{\BR}^\oplus\!\!\! \HHH (P) {\rm d}P,$$
 where
 $$H(P)=\sqrt{(P-P_{\rm f}-A(0))^2+M^2}+\hf.$$
For  every  fixed total momentum $P$, 
the existence of ground state of $\HHH (P)$ can be considered, but as far as we know there is no exact result on the existence of ground state of $H(P)$. 
See \cite{MiSp, HiSa, sasa13} for related results.

This paper is organized as follows. 
In Section 2 we set up notation and terminology, 
give the rigorous definition of $\HHH$ 
as a self-adjoint operator. 

Section 3 deals with localization and 
show that $\chi(\HHH)$ with smooth function 
$\chi$ with a support in $(-\infty, E+m)$ is compact. 

Section 4 establishes a HVZ-type theorem, 
i.e., $\sigma_{\rm ess}(\HHH)=[E+m, \infty)$, and proves that $\HHH$ has a ground state as a corollary of HVZ-type theorem.  

\section{Definitions and the main theorems}
In this section we define $\HHH $ as 
a self-adjoint operator on a Hilbert space, 
and give the main theorem. 
A particle Hamiltonian is given by 
the semi-relativistic Schr\"oding operator with a rest mass $M$:
\begin{eqnarray}
\sqrt{p^2+M^2}+V.
\end{eqnarray}
We shall introduce assumptions on $V$ later. 
We suppose that $M\geq 0$ throughout this paper unless otherwise stated.

Let $\fff=
\oplus_{n=0}^\infty 
\fff_n(W)=
\oplus_{n=0}^\infty \left[
\otimes_s^n W\right]$
be the Boson Fock space over
Hilbert space 
$W=\oplus^{d-1}\LR$, $d\geq 3$.
Here $\otimes_s^0 W=\oplus^{d-1}\mathbb{C}$.
Although the physically reasonable choice of the spatial dimension 
is $d=3$, we generalize it.
The creation operator and the annihilation operator in $\fff$ are denoted by 
$\add(f)$ and $a(f)$, $f\in W$, respectively. 
They 
are 
 linear in the test function $f$ and 
satisfy 
canonical commutation relations: 
$$[a(f),\add(g)]=(\bar f , g)_W,\qquad [a(f),a(g)]=0=[\add(f),\add(g)].$$
Here and in what follows 
the scalar product $(f,g)_{\mathscr K}$ 
on a Hilbert space $\mathscr K$ is linear in 
$g$ and anti-linear in $f$. 
Formally $a^\#(f)$  
is written  as 
$\d a^\#(f)=\sum_{r=1}^{d-1}
\int a^{\# r}(k) f_r(k)dk$ 
for 
$f=\oplus_{r=1}^{d-1}f_r\in W$. 
We introduce assumptions on the dispersion relation $\omega$. 
\begin{assumption}\label{omega}
$\omega\in C^1 (\BR; \RR)$, 
$\nabla\omega\in L^\infty(\BR)$, 
 $\d \inf_{k\in\BR } \omega(k)= m$ 
with some $m>0$ and 
$\d \lim_{|k|\to\infty}\omega(k)=\infty$.
\end{assumption}
The free field Hamiltonian $H_{\rm f}$ is given by the second quantization of 
the multiplication operator 
by $\oplus^{d-1}\omega$ on $W$. 
Thus formally it is defined by
\begin{eqnarray}
H_{\rm f}=\sum_{r=1}^{d-1}
\int \omega(k)a^{\dagger r}(k)a^r (k) dk.
\end{eqnarray}
Let $e^r(k)=(e^r_1(k),...,e^r_d(k))$ be 
$d$-dimensional polarization vectors, i.e., 
$e^r(k)\cdot e^s(k)=\delta _{rs}$ and $k\cdot e^r(k)=0$ for $k\in\BR\setminus\{0\}$ and  $r=1,...,d-1$. 
Let $\vp$ be an ultraviolet cutoff function, 
for which we introduce assumptions below.
\begin{assumption}\label{h1}
$
\omega\sqrt\omega\vp \in\LR$ 
and $\vp(k)=\ov{\vp(-k)}$.
\end{assumption}
From this assumption and $\inf_k\omega(k)=m>0$ we can see that 
$\vp/\sqrt\omega,\sqrt{\omega}\vp\in\LR$. 
We fix $\vp$ satisfying Assumption 
\ref{h1} throughout this paper. 
For  each $x\in\BR$ a quantized radiation field 
$A(x)=(A_1(x),...,A_d(x))$ 
is 
given by 
\begin{eqnarray}
A_\mu(x)=\frac{1}{\sqrt 2}
\sum_{r=1}^{d-1}\int e_\mu^r(k)
\left\{ \frac{\hat\varphi(k)  e^{-ik\cdot x}}{\sqrt{\omega(k)}} a^{\dagger r}(k) + \frac{\hat\varphi(-k)  e^{ik\cdot x}}{\sqrt{\omega(k)}} a^r(k) \right\}dk.
\end{eqnarray}
Then 
 $\vp(k)=\ov{\vp(-k)}$ implies that  $A_\mu(x)$ is essentially self-adjoint for each $x$. 
 We denote the self-adjoint extension by the same symbol $A_\mu(x)$. 
We identify $\hhh$ with 
$\int^\oplus_{\BR} \!\fff {\rm d}x$, 
and under this identification we 
define  the self-adjoint operator 
$A_\mu$ by $\int_{\BR}^\oplus\! A_\mu (x) {\rm d}x$.  

The first task is to define the  operator $\HHH$ in \kak{HHH} as a self-adjoint operator.
The square root of $(p\otimes\one-A)^2+M^2$, 
$\sqrt{(p\otimes\one-A)^2+M^2}$, is defined through the spectral measure associated with 
self-adjoint operator $(p\otimes\one-A)^2+M^2$. It is however not trivial to show the self-adjoitness of 
$(p\otimes\one-A)^2$. 
Let $N=d\Gamma(\one)$ be the number operator on $\fff$, i.e., 
$\d N=\sum_{r=1}^{d-1}\int 
a^{\dagger r}(k) a^r(k) dk$. 
Let $C^\infty(\one\otimes N)=
\cap_{n=1}^\infty 
D(\one\otimes N^n)$. 
\begin{prop}
\label{hiroshima}
Suppose Assumption \ref{h1}.
 Then 
$(p\otimes\one -A)^2$ is essentially self-adjoint on 
$D(p^2\otimes \one )\bigcap C^\infty(\one\otimes N)
$. 
\end{prop}
\proof See \cite[Lemma 7.53]{lhb11}. 
\qed

The closure of 
$(p\otimes\one -A)^2
\lceil_{D(p^2\otimes \one )
\bigcap C^\infty(\one\otimes N)}$ is denoted by 
$(p\otimes\one -A)^2$ in what follows.
Thus 
$\sqrt{(p\otimes \one -A)^2+M^2}$ is defined through the spectral measure 
of $(p\otimes\one -A)^2$.
\begin{df}
The SRPF Hamiltonian is defined by 
\begin{align}
\HHH =\sqrt{(p\otimes \one -A)^2+M^2}+V\otimes \one +\one\otimes \hf 
\end{align}
with 
the domain 
\begin{align}
D(\HHH )=
D(\sqrt{(p\otimes \one -A)^2+M^2})\cap 
D(V\otimes \one)\cap D(\one\otimes \hf ).
\end{align}
\end{df}
We do not write tensor notation $\otimes$ for 
notational convenience in what follows. 
Thus 
$\HHH $ can be simply written as 
\begin{align}
\HHH =\sqrt{(p -A)^2+M^2}+V+\hf .
\end{align}

\begin{assumption}\label{assumption2}
(1) $V$ is non-negative and satisfies that $\d\lim_{|x|\to\infty} V(x)=\infty$. 
(2)
$V$ is  twice differentiable, 
 and $\partial_\mu V,  
 \partial_\mu^2 V \in L^\infty(\BR)$
  for 
   $\mu=1,...,d$, 
and  $D(V)\subset D(|x|)$. 
\end{assumption}
\begin{lem}
Suppose Assumption \ref{assumption2}. 
Then  $p^2+V$ is self-adjoint on $D(p^2)\cap D(V)$, 
and essentially self-adjoint on $C_{\rm c}^\infty(\BR)$. 
\end{lem}
\proof
Since $V\in L_{\rm loc}^2(\BR)$, 
$p^2+V$ is essentially self-adjoint on 
$C_{\rm c}^\infty(\BR)$. Take an arbitrary vector $\Psi\in C_{\rm c}^\infty(\BR)$.
We have 
$\d
\| (p^2+V)\Psi\|^2=\| p^2\Psi\|+\| V\Psi\|^2 + 2\sum_{\mu=1}^{d}\Re(p_\mu^2\Psi,V\Psi)$.
For all $\epsilon>0$ there exists 
$C_\epsilon>0$ such that
\begin{align*}
2\Re(p_\mu^2\Psi,V\Psi)
&=2\Re\{(p_\mu\Psi,Vp_\mu \Psi)-(p_\mu\Psi,[V,p_\mu]\Psi)\} \\
&\geq -2 \| \partial_\mu V \| \|p_\mu\Psi\| \|\Psi\|\geq -\epsilon\|p_\mu\Psi\|^2 -C_\epsilon\|\Psi\|^2.
\end{align*}
Thus
$
\| p^2\Psi\|^2+\| V\Psi\| \leq C(\| (p^2+V)\Psi\|+\|\Psi\|)
$
follows with some constant $C>0$.
$p^2+V\lceil_{D(p^2)\cap D(V)}$ is closed, and then it is self-adjoint.
\qed

We can also show the self-adjointness of 
$\HHH$ under Assumption \ref{assumption2}. 
It is established in \cite{hir13} that 
$\HHH$ for $M>0$ is essentially self-adjoint 
on $D(|p|)\cap D(\hf)$ for 
external potential 
$V$ such that $D(V)\subset D(|p|)$ and 
$\|Vf\|\leq a\||p|f\|+b\|f\|$ for all $f\in D(|p|)$ with $0\leq a<1$ and $b\geq0$.
We can also show a stronger statement on the self-adjointness of $\HHH$.
This is established in \cite{hh13}. 
We set
\begin{eqnarray}
{\ms H}_{\rm fin}=C_{\rm c}^\infty(\BR)\hat\otimes \fff_\infty,
\end{eqnarray}
where 
$\hat\otimes$ denotes the algebraic tensor product and 
$$\fff_\infty=L.H.\{\Omega,\, a^\dagger (h_1)\cdots a^\dagger(h_n) \Omega | h_j \in C_{\rm c}^\infty (\BR), j=1,\cdots,n, n\geq 1\}.
$$
\begin{thm}
Suppose Assumptions \ref{omega}, \ref{h1} and 
\ref{assumption2}. 
Then (1) and (2) follow. 

(1)
Let $M\geq 0$. 
Then 
$H$ is self-adjoint on $D(|p|)\cap D(V)\cap D(\hf)$ and essentially self-adjoint on $\ms H_{\rm fin}$.

(2)
Fix an arbitrary $M_0>0$.
Then there exists a constant 
$C=C(M_0)>0$ such that 
for all $\Psi\in D({\HHH })$ and $0\leq M\leq M_0$, 
\eq{relative bound}
\| |p| \Psi \|^2 + \| V\Psi \|^2 +\| \hf \Psi \|^2  \leq C \| ({\HHH }+\one) \Psi \|^2.
\en
\end{thm}
\proof
See \cite[Lemma 2.9]{hh13}.
\qed

The ground state energy, $E$, of  $\HHH $ 
is the bottom of the spectrum of 
$\HHH $:
\begin{eqnarray}
E =\inf\sigma(\HHH ).
\end{eqnarray}
When $M=0$, we denote  $\HHH_0$ 
and $E_0$ for $\HHH $ and $E $, respectively.
The main results of this paper are as follows:
\begin{thm}[HVZ theorem for SRPF model]\label{main}
Suppose Assumptions \ref{omega}, \ref{h1}\\ and 
\ref{assumption2}.
Then 
$\sigma_{\rm ess}(\HHH )=[E +m,\infty)$
 for all $M\geq 0$.
\end{thm}
This theorem provides 
that 
$\HHH$ has a ground state for all $M\geq0$.
We summarize this in the corollary below.
\begin{cor}[Existence of the ground state] \label{hiroshima-1}
Suppose Assumptions \ref{omega}, \ref{h1}\\ and \ref{assumption2}.
Then 
$\HHH$ has the unique  ground state $\gr$ 
for all $M\geq0$, and \\
$$\|\gr(x)\|_\fff\leq Ce^{-c|x|}$$ with some constants $c$ and $C$. 
\end{cor}
\proof
By Theorem \ref{main} the lowest eigenvalue of $H$ is discrete. Then the ground state of 
$H$ exists. 
The uniqueness of the ground state is shown in \cite[Corollary 6.2]{hir13} and 
and spatial exponential decay of the 
ground state in  \cite[Theorem 5.12]{hir13}. 
\qed

\section{Localization}
The main result in this section is 
to estimate the asymptotic behaviour of 
a commutator, which is given in Lemma \ref{lem3.2}.  
\subsection{Commutator estimates}
We show a fundamental lemma. 
\begin{lem}\label{prop1}
Let $z\in \CC\setminus\RR$. Then 
$\d \lim_{M\downarrow 0}(\HHH -z)^{-1}=
(H_0-z)^{-1}$ in the uniform  topology.
In particular 
$\d \lim_{M\downarrow 0}\chi(\HHH)=
\chi(H_0)$ in the uniform  topology for all $\chi\in C_{\rm c}^\infty(\RR)$ and $\d \lim_{M\downarrow 0}E =E_0$.
\end{lem}
\proof
Let 
$\Psi\in \hhh$
and we set  $\Phi=(H-z)^{-1}\Psi$.
Let $E_\lambda$ be the spectral projection associated with the self-adjoint operator 
$|p-A|$. 
We have 
$$
\| (\HHH -z)^{-1}\Psi- (\HHH_0-z)^{-1} \Psi \|^2
\leq 
\frac{1}{|\Im z|^2}\int_0^\infty 
\!\!\! 
\left(\frac{M^2}{\lambda +\sqrt{\lambda^2+M^2}}\right)^2 d\| E_\lambda\Phi\|^2 
\leq 
\frac{M^2\|\Psi\|^2}{|\Im z|^2}.
$$
Then $\d \lim_{M\downarrow 0}(\HHH -z)^{-1}=(H_0-z)^{-1}$ is obtained, and  $E\to E_0$ follows. 
By the Helffer-Sj\"ostrand  formula \cite{HeSj} we have
\begin{eqnarray}
\chi(\HHH)=\frac{1}{2\pi i} \int_{\mathbb{C}}\!
\frac{\partial \tilde \chi(z)}{\partial{\bar{z}}}
(z-\HHH)^{-1} 
{\rm d}z {\rm d}\bar{z}.
\end{eqnarray}
Here 
${\rm d}z{\rm d}\bar{z}=-2i{\rm d}x{\rm d}y$, 
 $\d \frac{\partial}{\partial{\bar{z}}}
=\frac{1}{2}\left(
\frac{\partial}{\partial {x}}
+i\frac{\partial}{\partial {y}}\right)$ and
$\tilde{\chi}$ is an almost analytic extension of $\chi$, which satisfies that
\begin{align}
\label{h4}
&\tilde{\chi}(x)=\chi(x),\quad x\in\mathbb{R},\\
\label{h5}
&\tilde{\chi}\in C_{\mathrm{c}}^{\infty}(\mathbb{C}),\\
\label{h6}
&\left|\frac{\tilde{\chi}(z)}{\partial{\bar{z}}}\right|\leq
C_{n}|\Im z|^{n},\quad n\in\mathbb{N}.
\end{align}
Then 
$$
\|\chi(\HHH)-\chi(\HHH_0)\|\leq
\frac{1}{\pi} 
\int_{\mathbb{C}}\!
\left
\|
\frac{\partial \tilde \chi(z)}{\partial{\bar{z}}}
((z-\HHH)^{-1}-(z-\HHH_0)^{-1})
\right \| 
{\rm d}x {\rm d} y.
$$
We see that for all $z\in \supp{\tilde\chi}\setminus\mathbb{R}$,  
$
\Vert
\frac{\partial \tilde{\chi}(z)}{\partial{\bar{z}}}
 (z-\HHH)^{-1}\Vert\leq C_1|\Im z|$ 
and
$
\d\lim_{M\downarrow0}
 (z-\HHH)^{-1}= (z-\HHH_0)^{-1}$ uniformly.
Then 
by the Lebesgue dominated convergence theorem $\d\lim_{M\downarrow0}\chi(\HHH)=\chi(\HHH_0)$ is obtained.
\qed

\begin{lem}\label{coro}
Suppose Assumptions \ref{omega}, \ref{h1} 
and  \ref{assumption2}. 
Fix an arbitrary  $M_0>0$.
Then there exists a constant 
$C=C(M_0)>0$ such that 
for all $\Psi\in D({H})$ and $0\leq M\leq M_0$, 
$$
\|(N+\one )\Psi\|\leq \frac{C}{m} \left(\| \HHH \Psi\| + \| \Psi \|\right).
$$
\end{lem}
\proof
For all $\Psi\in D(\HHH )(\subset D(\hf ))$ 
we have
$\|N \Psi\|\leq 
\frac{1}{m}\| \hf \Psi\|$.
Then   
the corollary follows from 
the bound 
$
\| |p| \Psi \|^2 + \| V\Psi \|^2 +\| \hf \Psi \|^2  \leq C \| ({\HHH }+\one) \Psi \|^2$ 
shown in \kak{relative bound}.
 \qed

We shall divide the configuration space $W$ 
as $W=W_0\oplus W_\infty$, 
where $W_0$ denotes the set of functions supported on small momenta, 
and $W_\infty$ on large momenta. 
Since 
$\fff=\fff(W_0\oplus W_\infty)\cong \fff(W_0)\otimes\fff(W_\infty)$, 
we have 
$\hhh\cong (\LR\otimes\fff(W_0))\otimes \fff(W_\infty)$. 
Thus we  introduce the extended Hamiltonian 
$\EEE $ acting in the extended Hilbert space \eq{exthilbert}
\widehat{\hhh}=\hhh\otimes \fff 
\en by
\begin{eqnarray}
\EEE =\HHH \otimes \one
_{\fff }+\one_{\hhh}\otimes \hf .
\end{eqnarray}
Under Assumptions \ref{omega}, \ref{h1} and \ref{assumption2} we can also see that 
$\EEE$ is essentially self-adjoint on $D(\HHH\otimes\one_\fff)\cap D(\one_\hhh\otimes\hf)$. We denote the unique self-adjoint extension by 
the same symbol $\EEE$. 
We set $j=(j_{0},j_{\infty})\in 
C^{\infty}(\mathbb{R}^{d};\mathbb{R}_{+})\times C^{\infty}(\mathbb{R}^{d};\mathbb{R}_{+})$, where $j_{0}$ and $j_{\infty}$ satisfy that
\begin{eqnarray}
j_{0}(k)=\left\{
\begin{array}{rl}
1 & \mbox{ if $|k| \leq 1$}\\
0 & \mbox{ if $|k| \geq 2$}
\end{array}
\right.
\text{and}\quad
j_{0}^{2}(k)+j_{\infty}^{2}(k)=1.
\end{eqnarray}
We also define the bounded operator 
$\hat{j}_R:
W
\to 
W\oplus W$
for $R>0$ by
\begin{eqnarray}
\hat{j}_{R}f
=\hat{j}_{0,R}f
\oplus\hat{j}_{\infty,R}f=
j_{0}(\frac{-i}{R}\nabla_k)f
\oplus j_{\infty}(\frac{-i}{R}\nabla_k)f.
\end{eqnarray}
 Let us also define 
 the isometry 
 $\ir : \fff \rightarrow \fff \otimes \fff $ 
 by 
\begin{align}
\label{3.24}
\ir \Omega&=\Omega_{\fff \otimes \fff }, \\
\ir \prod_{i=1}^{n} a^{\dagger}(h_{i})
\Omega&=\prod_{i=1}^{n}(a_{0}^{\dagger}(\hat{j}_{0,R}h_{i})+ a_{\infty}^{\dagger}(\hat{j}_{\infty,R}h_{i}))\Omega_{\fff \otimes \fff },
\end{align}
where
$
a^{\dagger}_{0}(\hat{j}_{0,R} f)=
a^\dagger (\hat{j}_{0,R}f)\otimes \one$,
$
 a_{\infty}^{\dagger}(\hat{j}_{\infty,R}f)=\one \otimes a^{\dagger}(\hat{j}_{\infty,R}f)$ and 
 $\Omega_{\fff\times\fff}=\Omega\otimes\Omega$. 
Let $\chi\in C_{\rm{c}}^\infty(\mathbb{R})$ be 
such that 
 $\supp\chi\subset (-\infty,E +m)$. 
 We shall show that 
 $\chi(\HHH )$ is a compact operator. 
Note that 
$\ir^\ast\ir=\one$. 
Then the key identity 
is  
\eq{key}
\chi(\HHH )-\ir^\ast D_R=
\ir^\ast \chi(\EEE )\ir,
\en
where 
the remainder term is 
$D_R= 
\ir \chi (\HHH )-
\chi(\EEE )\ir$. 
Note that 
the first term of the right-hand side of \kak{key}, $\ir^\ast \chi (\EEE )\ir$, is compact. 
We shall show that 
the remainder term 
$\ir^\ast D_R$ uniformly converges to zero as $R\to\infty$. 
Hence we derive that $\chi(\HHH)$ is compact. So we estimate the commutator 
$\chi(\EEE )\ir -\ir \chi(\HHH )$ 
for an arbitrary $\chi\in C_{\rm{c}}(\RR)$.
\begin{lem} \label{lem3.2}
Suppose Assumptions \ref{omega}, \ref{h1} and  
\ref{assumption2} and $M>0$.
Let $\chi\in C_{\rm{c}}^{\infty}(\mathbb{R})$.
Then 
\begin{eqnarray}\label{A}
\lim_{R\to \infty}\left\| 
\chi(\EEE )\ir -\ir \chi(\HHH ) \right\|=0.
\end{eqnarray}
\end{lem}
We prepare  several lemmas to prove Lemma \ref{lem3.2}.
By the Helffer-Sj\"ostrand formula, we have
\begin{eqnarray}\label{lem3.2.1.0}
 \chi(\EEE )\ir -
\ir \chi(\HHH)
=\frac{1}{2\pi i}\!\!\int_{\mathbb{C}}\!
\frac{\partial \tilde{\chi}(z)}{\partial {\bar{z}}}
(z-\EEE )^{-1}(\EEE \ir -\ir \HHH )(z-H)^{-1} {\rm d}z {\rm d}\bar{z}.
\end{eqnarray}
Here $\tilde{\chi}$ satisfies (\ref{h4})-(\ref{h6}).
We  set 
$T=(p-A)^2+M^2$ and $\widehat T=
T\otimes\one_\fff.$
Note that
we have
\begin{align}
\label{mori}\EEE \ir -\ir \HHH  =
(\hf \otimes \one_{\fff }
+\one_{\hhh}\otimes \hf )
\ir -\ir \hf 
+\widehat T^\han \ir -\ir T^\han  .
\end{align}
Let 
$$B_R=\left\| 
\lk 
(\hf \otimes \one_{\fff }
+\one_{\hhh}\otimes \hf )
\ir -\ir \hf \rk (N+1)^{-1}\right\|.$$
By Assumption \ref{h1} the first two terms of the right-hand side of \kak{mori} 
 can be estimated as follows.
\begin{lem}\label{DG}
Suppose Assumptions \ref{omega}, \ref{h1} and  \ref{assumption2}. 
Then $\d \lim_{R\to\infty} B_R=0$, and 
for all $z\in\mathbb{C}\setminus \mathbb{R}$,
\begin{align}\label{lem3.2.1.1}
\left\| 
(\EEE-z)^{-1} \lk 
(\hf \otimes \one_{\fff }
+\one_{\hhh}\otimes \hf )
\ir -\ir \hf \rk (\HHH -z)^{-1}
\right\| 
\leq \frac{C}{m}  \left(1+ \frac{|z|+1}{|\Im z|}\right)^2  
B_R.
\end{align}
\end{lem}
\proof 
See e.g., \cite[Proof of Lemma 3.4]{DG99}. 
\qed\\

We set 
\eq{gm}
G_x(k)=\vp(k) e^{-ik\cdot x}/\sqrt{\omega(k)} \in L^2(\BR_k).
\en
Let $\#=0 $ or $\infty$. 
Then the inverse Fourier transform of 
$\hat j_{\#R} G_x$  is given by 
$j_\#(\cdot/R) \tvp(\cdot-x)$, where $\tvp$ is the inverse Fourier transform of 
$\vp/\sqrt{\omega}$. 
 $A_\#(x)$ denotes $A(x)$ with cutoff function $G_x$ replaced by 
 $\hat j_{R \#} G_x$, 
and we set 
$$
A_{0}=
 \int_{\RR^d}^\oplus\!\!\! 
A_0(x)  {\rm d}x 
,\quad 
 A_{\infty}=
 \int_{\RR^d}^\oplus\!\!\! 
  A_\infty(x)  {\rm d}x.
$$
$A_0\otimes\one_\fff$ and $\one_\fff\otimes A_\infty$ are self-adjoint operators in $\widehat \hhh$.
We set 
$$
\widehat S=(p-A_0\otimes\one -\one \otimes A_\infty)^2+M^2.
$$
Formally $A_0\to A$ and $A_\infty\to 0$ as $R\to \infty$, then $\widehat S\to \widehat T$ as $R\to \infty$. 
Let $\widehat N=N\otimes\one+\one\otimes N$ be the number operator on $\fff\otimes\fff$, and 
set $C^\infty(\widehat N)=\cap _{k=1}^\infty D(\widehat N^k)$.  
\begin{lem}
Suppose Assumptions \ref{omega}, \ref{h1} and \ref{assumption2}. Then $\widehat S$ is essentially self-adjoint on 
$D(p^2)\cap C^\infty(\widehat N)$.
\end{lem}
\proof
This follows from the similar method for the 
proof of the essential self-adjointness of $(p-A)^2$ in Proposition \ref{hiroshima}. 
Let $(B_t)_{t\geq 0}$ be the $d$-dimensional Brownian motion defined on the Wiener space, and $\Ebb^x[\cdots]$ denotes the expectation with respect to the Wiener measure starting at $x\in\BR$.  
Let 
 $${\rm K}_\#=\oplus_{i=1}^d \int_0^t  j_\#(\cdot/R)\tvp(\cdot-B_s) {\rm d}B_s^i,\quad \#=0,\infty.$$
Define the quadratic form 
$$Q:\hhh\times\hhh\ni (\Phi,\Psi)\mapsto 
\ix\left[
(\Phi(B_0),e^{-i\A_1({\rm K}_0)-i\A_2({\rm K}_\infty)}
\Psi(B_t))\right]\in\CC.$$
See Appendix for the detail of functional integrations. 
Then we can see that 
there exists a strongly continuous one-parameter semigroup $S_t$
such that $Q(\Phi,\Psi)=(\Phi, S_t\Psi)$ and 
furthermore 
the generator of $S_t$ (denoted by $K$) satisfies that $K\Psi=\widehat S\Psi$ for 
$\Psi\in D(p^2)\cap C^\infty(\widehat N)$. 
We can also see that 
$e^{-tK} D(p^2)\cap 
C^\infty(\widehat N)
\subset D(p^2)\cap C^\infty(\widehat N)$.  
Thus $D(p^2)\cap C^\infty(\widehat N)$ is invariant domain for $e^{-tK}$, and $K$ is essentially self-adjoint on 
$D(p^2)\cap C^\infty(\widehat N)$ and then 
so is $\widehat S$. 
\qed
 
We denote the self-adjoint extension of 
$\widehat S\lceil_{D(p^2)\cap C^\infty(\widehat N)}$ by the same symbol 
$\widehat S$ in what follows.

\begin{lem}
\label{h1int}
It follows that 
$
\ir T \subset \widehat S\ir
$, i.e., 
the intertwining property 
$
\ir T =\widehat S\ir
$
holds on $D(T)$. 
\end{lem}
\proof
Since the intertwining property 
$\ir e^{-i\A({\rm K})}=
e^{-i\A_1({\rm K}_0)-i\A_2({\rm K}_\infty)}\ir$ holds 
by the functional integration we see that 
\begin{align*}
(\ir^\ast \Phi, e^{-tT}\Psi)&=
\ix\left[(\ir^\ast \Phi(B_0), e^{-i\A({\rm K})}\Psi(B_t))\right]\\
&=
\ix\left[( \Phi(B_0), e^{-i\A_1({\rm K}_0)-i\A_2({\rm K}_\infty)}\ir \Psi(B_t))\right]=
(\Phi, e^{-t\widehat S}\ir\Psi),
\end{align*}
where $K=\oplus_{i=1}^d \int_0^t \j_s\tvp(\cdot-B_s) {\rm d}B_s^i$. 
Take the derivative at $t=0$ for $\Psi\in D(T)$. Then the lemma follows.
\qed

\begin{lem}
\label{h7}
It follows that 
\begin{align}
&\left\| (\EEE-z)^{-1} \left(\widehat T^\han\ir  - \ir T^\han  \right) (\HHH-z)^{-1}\Psi\right\|\non \\
&
\leq\frac{2}{\pi}
 \int_0^\infty \frac{dw}{\sqrt{w}} 
\|
 (\EEE -z)^{-1} \{(\widehat T+w)^{-1}\widehat T-(\widehat S+w)^{-1}\widehat S\} \ir (\HHH-z)^{-1}\Psi\|
 \label{integral}
\end{align}
for all $\Psi\in D(T)$ and $z\in\mathbb{C}\setminus\mathbb{R}$. 
\end{lem}
\proof
Using  
$ K^\han =(2/\pi)\int_0^\infty K(K+w)^{-1} /
\sqrt w {\rm d}w$ for strictly positive self-adjoint operator $K$ and 
$\widehat T\ir -\ir T=(\widehat T-\widehat S)\ir
$ 
on $D(T)$
by Lemma \ref{h1int}, we can derive (\ref{integral}).
\qed

We shall  estimate 
the integrand 
$\|
 (\EEE-z)^{-1} \{(\widehat T+w)^{-1}\widehat T-(\widehat S+w)^{-1}\widehat S\} \ir (\HHH -z)^{-1}\Psi\|$ 
 of (\ref{integral}).

\begin{lem}\label {h2bound}
 $\widehat T^{\han} (\EEE-z)^{-1}$,
 $\widehat S^{\han} (\EEE-z)^{-1}$ and $T (\HHH -z)^{-1}$ are bounded for all $z\in \mathbb{C}\setminus\mathbb{R}$.
\end{lem}
\proof 
For all $\Psi\in D(\EEE )$ we have
\begin{align}
\|\widehat T^{\han}\Psi\|
\leq
\sum_{\mu=1}^{d} \| p_\mu \Psi\|+ 
d\sqrt 2\left\|\frac{\hat\varphi}{\sqrt{\omega}}\right\|  \| (N+\one)^{\han}\Psi\|+M\|\Psi\|
\leq C\|(\EEE+\one)\Psi\|.
\end{align}
Then 
 $\widehat T^{\han} (\EEE-z)^{-1}$ is bounded. 
The boundedness  
of $\widehat S^{\han}(\EEE-z )^{-1}$ and $T (\HHH-z )^{-1}$ are similarly proven.
Then the lemma follows.
\qed

Next we estimate 
 $\widehat T^{\han}\ir (\HHH-z)^{-1}$.
\begin{lem}
 $\widehat T^{\han}\ir (\HHH-z)^{-1}$ is bounded, and there exists $C>0$ such that 
 \begin{eqnarray}
 \sup_{R>0} \|\widehat T^{\han}\ir (\HHH-z)^{-1}\|<C\left(1+\frac{1+|z|}{|\Im z|}\right)
 \end{eqnarray}
 for all $z\in\mathbb{C}\setminus\mathbb{R}$.
 \end{lem}
\proof
Since $\hhh_{\rm fin}$ is a core of $\HHH$, 
for all $\Psi\in D(\HHH)$ there exists a sequence $\{\Psi_{j}\}$ such that 
$\d \Psi_j\in \hhh_{\rm fin}$,
$\d \lim_{j\to\infty}\Psi_j=(\HHH -z)^{-1}\Psi$ and $\d \lim_{j\to\infty}H\Psi_j=H(\HHH-z)^{-1}\Psi$. 
Note that $\ir  \Psi_j\in D(\widehat T^{\han})$. For all $\Phi\in D(\widehat T^{\han})$ we have
\begin{align*}
&|(\widehat T^{\han}\Phi, \ir (\HHH-z)^{-1}\Psi)|
=\lim_{j\to\infty}|(\widehat T^{\han}\Phi, \ir \Psi_j)| \\
&\leq 
\lim_{j\to\infty} \| \Phi\|
\lk 
 \sum_{\mu=1}^d\| p_\mu\Psi_j\|+d\sqrt 2
 \left\|\frac{\hat\varphi}{\sqrt{\omega}}\right\| 
\| (N^{\han}\otimes\one +\one\otimes N^{\han})\ir \Psi_j\|+M\|\Psi_j\|
\rk  \\
&
=
\|\Phi\|\lk
 \sum_{\mu=1}^d\| p_\mu(\HHH-z)^{-1}\Psi\|+
 d\sqrt 2\left\|\frac{\hat\varphi}{\sqrt{\omega}}\right\| 
\| N^\han  
 (\HHH-z)^{-1}\Psi\|+M\|
 (\HHH-z)^{-1}\Psi\|\rk
\\
&\leq C\|\Phi\| \|(\HHH +\one) (\HHH-z)^{-1}\Psi\|=C\|\Phi\| \| \one + (\one-z)(\HHH-z)^{-1}\Psi\|\\
&\leq C\left(1+\frac{1+|z|}{|\Im z|}\right)\|\Phi\| \|\Psi\|,
\end{align*}
where we used
$(N^{\han}\otimes\one+\one\otimes N^{\han})\ir=\ir N^{\han}$. 
Thus $\widehat T^{\han}\ir(\HHH -z)^{-1}$ is bounded uniformly in $R$.
\qed

\begin{lem}
\label{h8}
For all $\epsilon>0$ and $z\in\mathbb{C}\setminus\mathbb{R}$ there exists $C_R$ such that 
$\d \lim_{R\to\infty}C_R=0$ and 
\begin{align}
&\| (\EEE-z)^{-1} (\widehat T^\han \ir  - \ir T^\han  ) (\HHH-z)^{-1}\|\leq 
(\epsilon+C_R)\left(1+\frac{1+|z|}{|\Im z|}\right)^2. \label{int}  
\end{align} 
\end{lem}
\proof
Since 
$\|(\widehat T+w)^{-1} \|\leq \frac{1}{M^2+w}$, 
$ \|(\widehat S+w)^{-1} \|\leq \frac{1}{M^2+w}$, 
the integrand of (\ref{integral}) can be estimated as
\begin{align}
&\|
 (\EEE-z)^{-1} \{(\widehat T+w)^{-1}\widehat T-(\widehat S+w)^{-1}\widehat S\} \ir (\HHH-z)^{-1}\Psi\|\non \\
& \leq 
\| \widehat T^{\han}(\EEE-z)^{-1}\|\, \|(\widehat T+w)^{-1} \|\, \| \widehat T^{\han}\ir 
(\HHH-z)^{-1}\| \non \\
& \quad+ \| \widehat S^{\han} 
(\EEE-z)^{-1}\|\, \| (\widehat S+w)^{-1}\|\, \|  \widehat S^{\han} \ir (\HHH-z)^{-1}\| \nonumber \\
&\leq \frac{C}{M^2+w}\left(1+\frac{1+|z|}{|\Im z|}\right)^2.
\end{align}
with some constant $C$ independent of $R$ and $M$.
Take an arbitrary $\epsilon>0$. Then there exists  a closed interval $[\delta, L]\subset (0,\infty)$ such that 
\begin{eqnarray}
\lefteqn{\| (\EEE-z)^{-1} (\widehat T^\han 
\ir  - \ir T^\han  )(\HHH-z)^{-1}\Psi\|} \nonumber\\
&\leq&\epsilon\left(1+\frac{1+|z|}{|\Im z|}\right)^2\|\Psi\| \non\\
&&\quad+\int_{[\delta, L]}\frac{dw}{\sqrt{w}} \| (\EEE-z)^{-1} \{(\widehat T+w)^{-1}\widehat T-(\widehat S+w)^{-1}\widehat S\} \ir (\HHH-z)^{-1}\Psi\| \nonumber \\
&=&\epsilon\left(1+\frac{1+|z|}{|\Im z|}\right)^2\|\Psi\|+\int_{[\delta, L]} \frac{dw}{\sqrt{w}} \{ 
\| (\EEE-z)^{-1} (\widehat T-\widehat S)(\widehat S+w)^{-1} \ir (\HHH-z)^{-1}\Psi\|\nonumber\\
&&\quad+
\|(\EEE-z)^{-1} (\one -w(\widehat T+w)^{-1})(\widehat S-\widehat T)(\widehat S+w)^{-1} \ir (\HHH-z)^{-1}\Psi\| \} 
.  \nonumber
\end{eqnarray} 
Note  that
$$
\widehat T-\widehat S= 
-2p\cdot(\xi-\eta) 
+(A\otimes\one_\fff) \cdot(\xi-\eta ) +(\xi-
\eta)\cdot (A_0\otimes\one_{\fff}+\eta),$$
where 
$\xi
=
A\otimes \one_\fff-A_0\otimes\one_\fff$ and 
$\eta=
\one_\fff\otimes  A_\infty$. 
It is shown in Subsection \ref{subsec} below that
\begin{align}
\label{340}
&\||p|(\widehat T+w)^{-1}(\EEE-z)^{-1}\|<C\left(1+\frac{1+|z|}{|\Im z|}\right),\\
\label{340'}
&\||A\otimes\one_\fff|(\widehat T+w)^{-1}(\EEE-z)^{-1}\|<C\left(1+\frac{1+|z|}{|\Im z|}\right),\\
\label{341}
&\| |\xi|(\widehat S+w)^{-1}\ir (\HHH-z)^{-1}\|<C_R\left(1+\frac{1+|z|}{|\Im z|}\right),\\
\label{342}
&\||\eta|(\widehat S+w)^{-1}\ir (\HHH-z)^{-1}\|<C_R\left(1+\frac{1+|z|}{|\Im z|}\right),\\
\label{344'}
&\||A_0\otimes\one_\fff|(\widehat S+w)^{-1}\ir (\HHH-z)^{-1}\|<C\left(1+\frac{1+|z|}{|\Im z|}\right),\\
\label{343}
&\||\xi|(\widehat T+w)^{-1} (\EEE-z)^{-1}\|<C_R\left(1+\frac{1+|z|}{|\Im z|}\right),\\
\label{344}
&\||\eta|(\widehat T+w)^{-1}(\EEE-z)^{-1}\|<C_R\left(1+\frac{1+|z|}{|\Im z|}\right).
\end{align}
\vspace{1mm}
Here $C_R$ is a constant such that  $\d \lim_{R\to\infty}C_R=0$. 
 Then we have
 \begin{align*}
& \| (\EEE-z)^{-1} 
(\widehat T-\widehat S)(\widehat S+w)^{-1} \ir (\HHH-z)^{-1}\|\\
&
=\|  (\EEE-z)^{-1}
 (
-2p\cdot(\xi-\eta) 
+\xi \cdot(\xi-\eta ) +
(\xi-\eta)\cdot \eta
)(\widehat S+w)^{-1} \ir (\HHH-z)^{-1}
\|\\
&\leq 
(2
\| |p|  (\EEE-z)^{-1} \| +\||\xi| (\EEE-z)^{-1}\|)
\cdot \|
|\xi-\eta|
(\widehat S+w)^{-1} \ir (\HHH-z)^{-1}\| \\
& +
\||\xi-\eta|(\EEE-z)^{-1}\|
\cdot 
\||\eta|
(\widehat S+w)^{-1} \ir (\HHH-z)^{-1}\|\\
&\leq a C_R\left(1+\frac{1+|z|}{|\Im z|}\right)^2
\end{align*}
with some constant $a$.  
Similarly we see that 
\begin{align*}
&
\| (\EEE-z)^{-1} 
(\one -w(\widehat T+w)^{-1})(\widehat S-\widehat T)(\widehat S+w)^{-1} \ir (\HHH-z)^{-1}\Psi\| \}\\
&
\leq 
(2
\| |p| (\one -w(\widehat T+w)^{-1}) (\EEE-z)^{-1} \| \\
& +\||\xi|(\one -w(\widehat T+w)^{-1}) (\EEE-z)^{-1}\|)
\cdot \|
|\xi-\eta|
(\widehat S+w)^{-1} \ir (\HHH-z)^{-1}\|\\
& +
\||\xi-\eta|(\one -w(\widehat T+w)^{-1})(\EEE-z)^{-1}\|
\cdot 
\||\eta|
(\widehat S+w)^{-1} \ir (\HHH-z)^{-1}\|\\
&\leq b(1+w)C_R\left(1+\frac{1+|z|}{|\Im z|}\right)^2
\end{align*}
with some constant $b$.
Together with them we obtain that 
\begin{align*}
&\| (\EEE-z)^{-1} (\widehat T^\han \ir  - \ir T^\han  ) (\HHH-z)^{-1}\Psi\| \\
& \leq 
\left(\epsilon+C_R
\int_{[\delta, L]} \frac{
a+(1+w)b}{\sqrt w} dw\right)\left(1+\frac{1+|z|}{|\Im z|}\right)^2\|\Psi\|.  
\end{align*} 
Since $C_R\to 0$ as $R\to \infty$, we obtain the lemma.
\qed

We give a proof of Lemma 
\ref{lem3.2}. 

{\it Proof of Lemma \ref{lem3.2}}:  
Let
$
c=2\int_{\mathbb{C}}\!
  |\frac{\partial \tilde{\chi}(z)}{\partial{\bar{z}}}|\left(1+\frac{1+|z|}{|\Im z|}\right)^2 
 {\rm d}x {\rm d}y$.
By (\ref{lem3.2.1.0}), Lemmas \ref{DG} and \ref{h8} we have
\begin{align}
&
\limsup_{R\to\infty} \Vert  \chi(\EEE )\ir -
\ir \chi(\HHH) \Vert
\leq c\lim_{R\to\infty}
 ( 
 B_R  + 
\epsilon+C_R)
=c\epsilon.
\end{align}
Since $\epsilon>0$ is arbitrary,
(\ref{A}) is obtained. Then the lemma follows. 
\qed

\subsection{Proof of \kak{340}-\kak{344}}\label{subsec}
It remains to show sequence of inequalities \kak{340}-\kak{344}. 
We prove these inequalities by functional integrations. 
Let $A_\#$ denote 
$A\otimes \one_{\fff}$ or $A_0\otimes\one_\fff+\one_\fff\otimes A_\infty$. 
The functional integration of the semigroup 
generated by $\half (p-A_\#)^2+M^2+w$ is given in Appendix. 
It can be then shown that 
\begin{align*}
&(\Phi, e^{-t(\half(p-A_\#)^2)}\Phi)_{\widehat\hhh}\\
&=
 \left\{
\begin{array}{ll}
\d 
\ix \left[(\Phi(B_0), 
e^{-i\A_1({\rm K}_0)-i\A_2({\rm K}_\infty)}\Psi(B_t))
\right],&
A_\#=
A_0\otimes\one_\fff+\one_\fff\otimes A_\infty,\\
\ \\
\d 
\ix \left[(\Phi(B_0), 
e^{-i\A_1({\rm K})}\Psi(B_t))
\right],&
A_\#=A\otimes \one_{\fff}
\end{array}
\right.
\end{align*}

\begin{lem}
\label{n}
Let 
 $M>0$. 
Then 
$(\widehat N+\one)
((p-A_\#)^2+M^2+w)^{-1}
(\widehat N+\one)^{-1}$ is bounded uniformly in $w\in [0,\infty)$.
\end{lem}
\proof 
We give the proof of the lemma in the case of $A_\#=A_0\otimes\one_\fff+\one_\fff\otimes A_\infty$. In another case, we can prove it in a similar manner.
Let $N_1=N\otimes\one_\fff$ and $N_2=\one_\fff\otimes N$. Then $\widehat N=N_1+N_2$.  
We see that 
$$
((N_1+N_2)\Phi, ((p-A_\#)^2+M^2+w)^{-1}\Psi)=
2\int_0^\infty 
e^{-t\half(M^2+w)}
((N_1+N_2)\Phi, e^{-t(p-A_\#)^2}\Psi) {\rm d}t.$$
We have 
\begin{align*}
&((N_1+N_2)\Phi, ((p-A_\#)^2+M^2+w)^{-1}\Psi)\\
&=
2\int_0^\infty 
\!\!\!
{\rm d}t e^{-t\half(M^2+w)}
\ix \left[
((N_1+N_2)\Phi(B_0), 
e^{-i \A_1({\rm K}_0)-i\A_2({\rm K}_\infty)} \Psi(B_t))\right]\\
&=2
\int_0^\infty 
\!\!\!\!
{\rm d}t e^{-t\half(M^2+w)}
\ix \!\left[\!
(\Phi(B_0), 
e^{-i \A_1({\rm K}_0)-i\A_2({\rm K}_\infty)} 
\sum_{j=1,2}
e^{i \A_j({\rm K}_j)}
N_j
e^{-i \A_j({\rm K}_j)}
\Psi(B_t))\!\right]\!,
\end{align*}
where ${\rm K}_1={\rm K}_0$ and ${\rm K}_2={\rm K}_\infty$. 
We have 
$$e^{i \A_j({\rm K}_j)}
N_j
e^{-i \A_j({\rm K}_j)}
=N_j-i\Pi({\rm K}_j)-\half {\rm q}({\rm K}_j,{\rm K}_j).$$
Here $\Pi_j({\rm K}_j)=i[N_j, \A_j({\rm K}_j)]$. 
We know that 
$\|\Pi_j({\rm K}_j)\Phi\|\leq \|{\rm K}_j\|\|(N_j+\one)^{\han}\Phi\|$ and ${\rm q}({\rm K}_j,{\rm K}_j)\leq \|{\rm K}_j\|^2$. 
Hence by Lemma \ref{hiroshima1} below, 
there exist constants $c_1$ and $c_2$ such that 
\begin{align*}
\left|
\ix \left[
(\Phi(B_0), \Pi_j({\rm K}_j)\Psi(B_t))\right]\right|
\leq 
c_j \|\Phi\|
\|(N_j+\one)^\han\Psi\|
\sqrt t, \quad j=1,2.
\end{align*}
We can also see that 
$|(\Phi(B_0), \half {\rm q}({\rm K}_j, {\rm K}_j)
\Psi(B_t))|\leq 
\|\Phi(x)\| \|{\rm K}_j\|^2\Psi(B_t)\|$.
Hence by Lemma \ref{hiroshima1} again there exist constants $d_j$ such that 
\begin{align*}
\left|
\ix \left[
(\Phi(B_0), \half {\rm q}({\rm K}_j, {\rm K}_j)
\Psi(B_t))\right]\right|
&\leq 
\int_{\BR} \!\!\! {\rm d}x 
\|\Phi(x)\|
\Ebb^x[\|{\rm K}_j\|^4]^\han 
\Ebb^x[\|\Psi(B_t)\|^2]^\han\\
&\leq 
d_j
\|\vp/\sqrt\omega\|^2 
\|\Phi\|
\|\Psi\|
 t,\quad j=1,2. 
\end{align*}
Then we have 
\begin{align*}
&|(\widehat N\Phi, ((p-A_\#)^2+M^2+w)^{-1}\Psi)|\\
&=
2\int_0^\infty 
e^{-t\half(M^2+w)}
{\rm d}t
\|\Phi\| (\sqrt t \sum_{j=1,2}c_j
\|(N_j+\one)^\han\Psi\|+td_j\|\Psi\|)
\leq C \|\Phi\|\|(\widehat N+\one)\Psi\|
\end{align*}
with some constant $C$ independent of $w$. 
Hence $(\widehat N+\one)
((p-A_\#)^2+M^2+w)^{-1}(\widehat N+\one)^{-1}$ is bounded uniformly in $w$. 
\qed

\begin{lem}\label{hiroshima1}
There exist constants $c_1$ and $c_2$ 
such that 
$\Ebb^x[\|{\rm K}_j\|^2]\leq tc_1\|\vp/\sqrt\omega\|^2$ and 
$\Ebb^x[\|{\rm K}_j\|^4]\leq t^2c_2\|\vp/\sqrt\omega\|^{4}$.
\end{lem}
\proof
See \cite[Theorem 4.6]{hir00} and \cite[Lemma 7.21]{lhb11}.
\qed
\begin{lem}\label{hidaka}
Let $0<M$ and $\chi\in C_{\rm c}^\infty(\BR)$.
Then for all $w\in[0,\infty)$
\begin{eqnarray}
\| |p|(|p-A_\#|^2+M^2+w)^{-1}(\EEE-z)^{-1}\|&\leq& C_M\left(1+\frac{1+|z|}{|\Im z|}\right), \label{hi-1} \\
\hspace{-5mm}\| |A\otimes\one_\fff|(|p-A_\#|^2+M^2+w)^{-1}(\EEE-z)^{-1}\|&\leq& C_M\left(1+\frac{1+|z|}{|\Im z|}\right), \label{hi-1'} \\
\| |A_R|(|p-A_\#|^2+M^2+w)^{-1}(\EEE-z)^{-1}\| &\leq& C_RC_M\left(1+\frac{1+|z|}{|\Im z|}\right), \label{hi-2}\\
\| |p|(|p-A_\#|^2+M^2+w)^{-1}\ir(\HHH-z)^{-1}\| &\leq& C_M\left(1+\frac{1+|z|}{|\Im z|}\right), \label{hi-3}\\
\hspace{-5mm}\| |A_0\otimes \one_\fff|(|p-A_\#|^2+M^2+w)^{-1}\ir(\HHH-z)^{-1}\| &\leq& C_M\left(1+\frac{1+|z|}{|\Im z|}\right), \label{hi-3'}\\
\| |A_R|(|p-A_\#|^2+M^2+w)^{-1}\ir(\HHH-z)^{-1}\| &\leq& C_RC_M\left(1+\frac{1+|z|}{|\Im z|}\right) \label{hi-4}
\end{eqnarray} 
with $C_{R}\in o(R^0)$ $(R\to\infty)$ and some positive constant $C_M$.
\end{lem}
\proof
First we prove (\ref{hi-1}).
For fixed $A_\#=A\otimes\one$ or $A_0\otimes\one+\one\otimes A_\infty$ we have
\begin{align}
& \| |p| (|p-A_\#|^2+M^2+w)^{-1}(\EEE-z)^{-1} \|^2 \non \\
& \leq 2 \{\| |p-A_{\#}| (|p-A_\#|^2+M^2+w)^{-1}(\EEE-z)^{-1}\|^2 \non\\
& +\| |A_{\#}|(|p-A_\#|^2+M^2+w)^{-1}(\EEE-z)^{-1}\|^2 \}. \label{hida-1}
\end{align}
By Lemma \ref{hiroshima} we have
\begin{align}\label{hida-2}
&\| |A_{\#}|( |p-A_{\#}|^2+M^2+w)^{-1}(\EEE -z)^{-1} \| \non\\
&\leq 2 \|G\| \|(N+1)( |p-A_{\#}|^2+M^2+w)^{-1}(N+1)^{-1}\| \|(N+1)(\EEE -z)^{-1}\|\non\\
&\leq C_M'\left(1+\frac{1+|z|}{|\Im z|}\right)
\end{align}
with some constant $C_M'$. Together with (\ref{hida-1}) and (\ref{hida-2}) we obtain that
\begin{align}
&\| |p| (|p-A_\#|^2+M^2+w)^{-1}(\EEE-z)^{-1} \| \leq \sqrt{2}\left(\frac{1}{\sqrt{M^2+w}}+C_M'\right)\left(1+\frac{1+|z|}{|\Im z|}\right).
\end{align}
Then (\ref{hi-1}) is obtained.
Next we prove (\ref{hi-2}). 
For  an arbitrary $\epsilon>0$
there 
 exists a normalized vector $\Psi_R$ 
 such that
\begin{eqnarray}
\| A_{R,\mu}(\widehat N+\one)^{-\han} \|
\leq \| A_{R,\mu}(\widehat N+\one)^{-\han}\Psi_R \|+\frac{\epsilon}{3}.
\end{eqnarray}
Let $G_x\in \LR$ be in \kak{gm}. 
Notice that 
$\| G_x\|$ is independent of $x$.
There exists $L>0$ such that 
$\Psi_{R,L}=\chi_{\{|x|\leq L\}} \Psi_R$ satisfies that 
$\| \Psi_R -\Psi_{R,L} \|_\hhh \leq \frac{\epsilon}{3(\| G_x \|+1)}$. 
Then we have
\begin{eqnarray}\label{dini}
\| A_{R,\mu}(\widehat N+\one)^{-\han} \|
\leq \| \| j_{R}(-i\nabla)G \|\Psi_{R,L} \|+\frac{2\epsilon}{3}.
\end{eqnarray}
Here $j_{R}$ stands for $\one-j_{0}(k/R)$ 
or $j_\infty(k/R)$.
\begin{eqnarray}
\| \| j_{R}(-i\nabla)G \|_{W}\Psi_{R,L} \|_\hhh^2
&=& \int_{|x|\leq L} \left\| \| j_{R}\hat{G}_x \|_{W} \Psi_R(x)\right\|_\fff^2 {\rm d}x \nonumber \\
&\leq& \int_{|x|\leq L} \left\| 
\| \chi_{\{|k|\geq R\}}\hat{G}_x \|_{W} 
\Psi_R(x)\right\|_\fff^2{\rm d}x.
\end{eqnarray}
Let 
$f_R(x)=\| \chi_{\{|k|\geq R\}}\hat{G}_x \|_W$. 
$f_R$ is continuous, and for each 
$x\in\BR$,  
it monotonically converges to $0$ as 
$R\to\infty$.
Then $f_R$ converges to $0$ uniformly on 
any compact set by Dini's theorem.
Thus we see that
$\d \lim_{R\to\infty}\sup_{|x|\leq L}\| 
\chi_{\{|k|\geq R\}}\hat{G}_x \|=0.$ 
Since $\Psi_R$ is normalized,
the right-hand side of (\ref{dini}) converges to $0$ as $R\to\infty$.
Thus there exists some $R_0>0$ such that for all $R>R_0$,
\begin{eqnarray}\label{dini2}
\| \| j_{R}(-i\nabla)G \|_{W}\Psi_{R,L} \|_\hhh^2<\frac{\epsilon}{3}.
\end{eqnarray}
By (\ref{dini}) and (\ref{dini2}) we see that
$
\| A_{R,\mu}(\widehat N+\one)^{-\han} \|<\epsilon
$
for all $R>R_0$. Then 
\begin{eqnarray}
\lim_{R\to\infty}\| A_{R,\mu}(\widehat N+\one)^{-\han} \|=0.
\end{eqnarray}
We have
$
\|(\widehat N+\one) (|p-A_{\#}|^2+M^2+w)^{-1} (\widehat N+\one)^{-1} \|\leq C_M
$ 
in Lemma \ref{n}.
Thus (\ref{hi-2}) is obtained.
(\ref{hi-1'}) and (\ref{hi-3})-(\ref{hi-4}) are also shown in a similar way.
\qed


\section{HVZ-type theorem}
\begin{lem}\label{main1}
For all $M\geq 0$ it  follows that 
$\sigma_{\rm ess}(\HHH )\subset [E +m, \infty).$
\end{lem}
\proof
Let $\chi\in C_{\rm{c}}(\mathbb{R})$ 
be such that 
$\supp \chi \subset (-\infty, E +m)$.
It suffices to show that $\chi(\HHH)$ is compact for all positive $M>0$.
Actually $\lim_{M\to+0}\chi(\HHH)=\chi(\HHH_0)$ in the uniform topology by Lemma~\ref{prop1}.
Let $P_{0}$ be the projection from 
$\fff$ to the subspace spanned by $\Omega$, i.e., $P_{0}\Psi=(\Omega,\Psi)\Omega$.
Since $\supp\chi \subset (-\infty, E +m)$ and 
$\sigma(\hf)=\{0\}\cup[m,\infty)$, 
we see that $\one_{\hhh}\otimes P_{0}$ leaves $\chi(\EEE )$ invariant: 
\begin{eqnarray}
\chi(\EEE )=(\one_{\hhh}\otimes P_{0})\chi(\EEE ). \label{lem3.3.1}
\end{eqnarray}
We also see that
\begin{eqnarray}
\ir ^{*}(\one_{\hhh}\otimes P_{0})\ir =\Gamma(\hat{j}_{0,R}^{2}). \label{lem3.3.2}
\end{eqnarray}
By Lemma \ref{lem3.2} and (\ref{lem3.3.1})
we have 
\begin{eqnarray}
\chi(\HHH )=\ir ^{*}\ir \chi(\HHH )=\ir ^{*}\chi(\EEE )\ir+o(R^{0})=\ir ^{*}(\one_{\hhh}\otimes P_{0})\chi(\EEE )\ir+o(R^{0}).
\end{eqnarray}
Here $o(R^{0})$ converges to $0$ as $R\to\infty$ in the uniform norm.
By (\ref{lem3.3.2}) and Lemma \ref{lem3.2} again we have
\begin{eqnarray}
\chi(\HHH )=\Gamma(\hat{j}_{0,R}^{2})\chi(\HHH )+o(R^{0}).
\end{eqnarray}
Then we can see that
\begin{eqnarray}\label{lem3.3.3}
\chi(\HHH )=\sum_{l=0}^{L}\Gamma(\hat{j}_{0,R}^{2})\one_{\{l\}}(N)\chi(\HHH )+\Gamma(\hat{j}_{0,R}^{2})\one_{[L+1,\infty)}(N)\chi(\HHH )+o(R^{0}).
\end{eqnarray}
Since  $\|\Gamma(\hat{j}_{0,R}^{2})\|\leq 1$ and $\| \one_{[L+1,\infty)}(N)\Psi\| \leq (L+1)^{-1}\| N\Psi\|$ for $\Psi\in D(N)$, we see that
\begin{eqnarray}\label{lem3.3.4}
\| \Gamma(\hat{j}_{0,R}^{2})\one_{[L+1,\infty)}(N)\chi(\HHH) \|
\leq
 \frac{1}{L+1} \| N\chi(\HHH ) \|
\to 0
\end{eqnarray}
as 
$L\to\infty$. 
By (\ref{lem3.3.3}) and (\ref{lem3.3.4}) we have
\begin{eqnarray}
\chi(\HHH )=
\sum_{l=0}^{L}\one_{\{l\}}(N)\Gamma(\hat{j}_{0,R}^{2})\chi(\HHH )+o(L^{0})+o(R^{0}),
\end{eqnarray}
where $o(L^{0})$ converges to $0$ as $L\to \infty$ in the uniform norm.
Thus it suffices to show that  $\one_{\{l\}}(N)\Gamma(\hat{j}_{0,R}^{2})\chi(\HHH)$ is compact for each $l=0,1,2,\cdots$.
We obtain that
\begin{eqnarray}
\one_{\{l\}}(N)\Gamma(\hat{j}_{0,R}^{2})\chi(\HHH )=UB,
\end{eqnarray}
where
$B=(p^2+V)^{1/4} 
(\hf+\one)^{1/4}
\chi(\HHH )$ and 
$U=(p^2+V)^{-1/4}
\one_{\{l\}}(N)\Gamma(\hat{j}^{2}_{0,R})
(\hf +\one)^{-1/4}$ is a compact operator.
Since
$\| (p^2+V)^{\han}\Psi\|^2
\leq C \|(H+\one)\Psi\|^2$, 
$B$ is bounded. 
Thus 
$\one_{\{l\}}(N)
\Gamma(\hat{j}_{0,R}^{2})
\chi(\HHH )$ is compact.
Then it yields that 
\eq{chi}
\chi(\HHH)=
\lim_{R,L\to\infty} 
\left\{\sum_{l=0}^{L}\one_{\{l\}}(N)\Gamma(\hat{j}_{0,R}^{2})\chi(\HHH )\right\},
\en
which implies that 
$\chi(\HHH)$ is the limit of compact operators in the uniform topology.  Then 
$\chi(\HHH)$ is also compact. 
Hence $(-\infty, E+m)\cap \sigma(\HHH)$ is 
discrete spectrum. Then the lemma follows. \qed

{\it The proof of Corollary \ref{hiroshima-1}:} 
Since $[E, E+m)\cap \sigma (\HHH)$ is discrete by 
Lemma \ref{main1}, $\HHH$ has a ground state.  
The uniqueness and the exponential decay 
for the ground state are  shown in \cite{hir13}. 
\qed

\begin{lem}
\label{main2}
Suppose Assumptions 
\ref{omega}, \ref{h1} and \ref{assumption2}.
Then
$\sigma_{\rm ess}(\HHH )\supset [E +m, \infty).$
\end{lem}
\proof
First we  assume that $M>0$. 
Let $\Phi_M$ be a normalized ground state of $\HHH $. 
Take $\lambda\in (E +m,\infty)$ and 
$k_0=k_0(M,\lambda)\in\BR$ such that $\omega(k_0)=\lambda-E$. 
Let $h\in C_{\rm c}^\infty(\BR )$ be  
such that  $\|h\|=1$.
Set $h_n(k)=n^{d/2}h(n(k-k_0))e^{in^2
(k_0- k)}$.
Then 
$\| h_n\|=1$, 
$\d \text{w-}\lim_{n\to\infty}h_n=0$ and 
$\d \lim_{n\to\infty}\| (\omega-\omega(k_0))h_n \|=0$. 
Note that 
$\Phi_M\in D(\HHH)\subset D(\hf)\subset D(N)\subset D(\add(f))$. 
Set
$\tilde{h}_n=\oplus^{d-1}h_n\in W$ 
and 
$\Psi_n=a^\dagger(\tilde{h}_n)\Phi_M$. 
It holds that 
$\d \lim_{n\to\infty}\| \Psi_n\|=1$ and 
$\d \text{w-}\lim_{n\to\infty}\Psi_n=0$. 
We see that for $\Phi\in \hhh_{\rm fin}$
\begin{align}\label{3.55}
((\HHH -\lambda)\Phi,\Psi_n)
=
(\Phi,a^\dagger((\omega-\omega(k_0))\tilde{h}_n)\Phi_M) + 
([a^\dagger(\tilde{h}_n),T^\han  ]\Phi, \Phi_M).
\end{align}
Then 
\begin{eqnarray}\label{3.56}
\text{s-}\lim_{n\to\infty}a^\dagger((\omega-\omega(k_0))\tilde{h}_n)\Phi_M = 0.
\end{eqnarray}
Let us consider the commutator $[a^\dagger(\tilde{h}_n),T^\han ]$. We see that
\begin{eqnarray}\label{T.1}
[a^\dagger(\tilde{h}_n),T^\han  ]
 =\frac{2}{\pi}\int_0^\infty \frac{dw}{\sqrt{w}}[T(T+w)^{-1},a^\dagger(\tilde{h}_n)].
\end{eqnarray}
Let $G_\mu=\oplus_{r=1}^{d-1}G_\mu^r$, and $\d G_\mu^r(k)=G_\mu^r(k,x)=\frac{\hat\varphi(k) e_\mu^r(k) e^{-ik\cdot x}}{\sqrt{2\omega(k)}}$.
Then we have
\begin{align}\label{T.2}
& [T(T+w)^{-1},a^\dagger(\tilde{h}_n)] \non\\
& =-w(T+w)^{-1}[T,a^\dagger(\tilde{h}_n)](T+w)^{-1} \non\\
& =-w(T+w)^{-1}\{(p+A)\cdot[A,a^\dagger(\tilde{h}_n)]+[A,a^\dagger(\tilde{h}_n)]\cdot(p+A)\}(T+w)^{-1} \non\\
& =-w(T+w)^{-1}\{(p+A)\cdot (G,\tilde{h}_n) +(G,\tilde{h}_n)\cdot(p+A)\}(T+w)^{-1} \non\\
& =-2w(T+w)^{-1} (G,\tilde{h}_n) (T+w)^{-1}\cdot (p+A)-w(T+w)^{-1}\sum_{\mu=1}^d (i\partial_{x_\mu} G_\mu, \tilde{h}_n)(T+w)^{-1}. \non\\
\end{align}
Since $\Phi_M$ is a ground state, by (\ref{T.1}), (\ref{T.2}) and Lemma \ref{relative bound} we obtain that
\begin{eqnarray}\label{3.57-hiro}
|([a^\dagger(\tilde{h}_n),T^\han  ]\Phi, \Phi_M)|
=C \sum_{\mu=1}^{d}
\sup_{x\in\mathbb{R}^d}\left( |(G_\mu,\tilde{h}_n)| + 
|(\partial_{x_\mu}G_\mu,\tilde{h}_n)| \right)
\|\Phi\|\|\Phi_M\|.
\end{eqnarray}
Let $K_n=
([a^\dagger(\tilde{h}_n),T^\han  ]\lceil_{\mathcal{H}_{\rm fin}})^\ast
$.
\kak{3.57-hiro} implies that 
$\Phi_M\in D(K_n)$ and $\d\lim_{n\to\infty}K_n\Phi_M=0$.
Then we see that 
$\|(\HHH -\lambda)\Psi_n\|
\leq \|\add((\omega-\omega(k_0))\tilde h_n)\Phi_M\|+\|K_n\Phi_M\|$
by (\ref{3.55})-(\ref{3.57-hiro}), and that 
$$
\text{s-}\lim_{n\to\infty} (\HHH -\lambda)\Psi_n=0.
$$
Since  $\d \lim_{n\to\infty}\|\Psi_n\|= \|\Phi_M\|+\lim_{n\to\infty}\|a(h_n)\Phi_M\|=1$, 
the normalized vector $\tilde \Psi_n=\Psi_n/\|\Psi_n\|$ satisfies that 
$$
\text{s-}\lim_{n\to\infty} (\HHH -\lambda)\tilde \Psi_n=0.$$
Then 
$\{\tilde{\Psi}_n\}$ is a Weyl sequence for $\lambda$ and then 
 we obtain that $\lambda\in \sigma_{\rm ess}(\HHH )$ when $M>0$.

Next we assume that  $M=0$. 
In order to emphasize the dependence on $M$ we use $\HHH_M$ and $E_M$ 
for $\HHH$ and $E$, respectively.  
Since $\HHH_M $  converges to $\HHH_0$ in the uniformly resolvent sense. Then 
$E_M \to E_0$ as $M\to 0$.
Fix $\lambda\in (E+m,\infty)$.
Let $\{M_j\}_j$ be a sequence such that 
$M_j\to 0$ as $j\to \infty$. 
Suppose that $\lambda>E_j+m$ for all $j$. 
For each $M_j$, by the discussion  mentioned above for the case of $M>0$ 
there exist $n_j=n_j(M_j)$ such that 
\begin{align*}
&\|\add (\tilde h_{n_j})\Phi_{M_j}\|\leq 1+1/j,\\
&
|(\Phi, \add (\tilde h_{n_j})\Phi_{M_j})|\leq 1/j,\\
&
\|(H_{M_j}-\lambda) 
\add (\tilde h_{n_j})\Phi_{M_j}\|\leq 1/j.
\end{align*}
Set $Q_j=\add (\tilde h_{n_j})\Phi_{M_j}$.
Then 
$\d \lim_{j\to\infty} \|Q_j\|\to 1$ and 
$$\|(\HHH_0-\lambda)Q_j\|\leq 
\|(\HHH_0-\HHH_{M_j})Q_j\|+
\|(\HHH_{M_j}-\lambda)Q_j\|\leq \sqrt{M_j}(1+1/j)+1/j.$$
Let 
$\tilde{Q}_j=Q_j/\|Q_j\|$. 
Hence we conclude that 
$\{\tilde{Q}_j\}$ 
is a Weyl sequence for $\lambda$,  
and thus 
$\lambda\in \sigma_{\rm ess}(\HHH_0)$ follows. Then the lemma follows. 
\qed

{\it The proof of Theorem \ref{main}:}

The theorem follows from Lemmas \ref{main1} and \ref{main2}. 
\qed

\section{Appendix}
In this appendix we review  functional integral representations of the semigroup generated by 
models related to the Pauli-Fierz model. 
These representations play an important roles in this paper. The functional integral representation for the semigroup generated by the Pauli-Fierz model has been established in \cite{hir97}. By a minor modification  we can  also construct functional integral representations for models investigated in this paper. 

\subsection{Pauli-Fierz model}
The Feynman-Kac formula yields 
the path integral representation of the Schr\"odinger operator $\half p^2+V$ by 
\eq{fy}
(f, e^{-t(\half p^2+V)}g)=
\ix \left[e^{-\int_0^tV(B_s) {\rm d}s} 
\ov{f(B_0)}g(B_t)\right].
\en
On the other hand the  Pauli-Fierz  model is defined by the minimal coupling of $p^2/2+V+\hf$ 
with a quantized radiation field $A(f)$ by  
$$H_{PF}=\half(p-A(f))^2+V+\hf$$
as a linear operator in $\hhh=\LR\otimes\fff(W)$, 
where $A(f)=(A_1(f),\cdots,A_d(f))$ describes quantized radiation field with cutoff function 
$f$ such that $f/\sqrt \omega\in\LR$, i.e, 
$A_\mu(f)=\int^\oplus_{\BR} A_\mu(f,x) {\rm d}x$ and 
\begin{eqnarray}
A_\mu(f,x)=\frac{1}{\sqrt 2}
\sum_{r=1}^{d-1}\int e_\mu^r(k)
\left\{ \frac{\hat f(k)  e^{-ik\cdot x}}{\sqrt{\omega(k)}} a^{\dagger r}(k) + 
\frac{\hat f(-k)  e^{ik\cdot x}}{\sqrt{\omega(k)}} a^r(k) \right\}dk.
\end{eqnarray}
We can give the functional integral representation of 
$e^{-tH_{PF}}$ in \cite{hir97}.
Let
$${\rm q}(F, G)=\half \sum_{\mu,\nu=1}^d (\hat F_\mu, 
\delta_{\mu\nu}^\perp\hat G_\nu)$$ be the quadratic form 
on $\oplus^d \LR$,
where
$\delta_{\mu\nu}^\perp(k)=\delta_{\mu\nu}-k_\mu 
k_\nu/|k|^2$ denotes the transversal delta function.  
Let $\A(F)$ be a 
Gaussian random variables on a probability space $(Q,\Sigma,\mu)$, which  is indexed by 
$F=(F_1,\cdots,F_d)\in \oplus^d \LR$. 
The  mean of $\A(F)$ is 
zero and the covariance is given by 
$\Ebb[\A(F)\A(G)]= {\rm q}(F, G)$. 
Furthermore 
we introduce the Euclidean version of $\A$.
Let
\eq{t1}
{\rm q}_E(F, G)= 
\half \sum_{\mu,\nu=1}^d 
(\hat F_\mu, 
\delta_{\mu\nu}^\perp\hat G_\nu)
\en
be the quadratic form on 
$\oplus^d L^2(\RR^{d+1})$.
On the right-hand side of \kak{t1}, we  note  that 
$(\hat F_\mu, 
\delta_{\mu\nu}^\perp\hat G_\nu)=
\int _{\RR\times \RR^d}
\ov{\hat F_\mu}(k_0,k)
\delta_{\mu\nu}^\perp(k) 
\hat G_\nu(k_0, k) {\rm d}k_ 0{\rm d}k
$. 
Let $\AA(F)$ be a Gaussian random variables on a probability space $(Q_E,\Sigma_E,\mu_E)$, which  is indexed by 
$F\in \oplus ^d L^2(\RR^{d+1})$.
The  mean of $\AA(F)$ is 
zero and the covariance is given by 
$
\Ebb[\AA(F)\AA(G)]=
{\rm q}_E(F,G)$. 
Let us identify $\hhh$ with $L^2(\BR;\fff)$. Thus 
$\Phi\in\hhh$ can be an $\fff$-valued $L^2$-function on $\BR$, 
$\BR\ni x\mapsto \Phi(x)\in\fff$. 
It is well known that 
there exists the family of isometries 
$J_t: L^2(Q)\to L^2(Q_E)$ ($t\in\RR$) 
and $\j_t:L^2(\BR)\to L^2(\RR^{d+1})$
($t\in\RR$) such that 
$J_t^\ast J_s=e^{-|t-s|\hf}$ and 
$\j_t^\ast \j_s=e^{-|t-s|\omega(-i\nabla)}$. 
By the Feynman-Kac formula \kak{fy} 
it is straightforward to see that
\eq{hh12}
(\Phi, e^{-t(\half p^2+V+\hf)}\Psi)=
\ix \left[e^{-\int_0^tV(B_s) {\rm d}s} 
(J_0\Phi(B_0), J_t\Phi(B_t))_{L^2(Q_E)}\right].
\en
Adding an interaction we  also see 
that 
\eq{fy2}
(\Phi, e^{-tH_{PF}}\Psi)=
\ix \left[e^{-\int_0^tV(B_s) {\rm d}s} 
(J_0\Phi(B_0), e^{-i\AA({\rm K}_E)}J_t\Phi(B_t))_{L^2(Q_E)}\right].
\en
Here 
$${\rm K}_E=\oplus_{i=1}^{d}
\int _0^t \j_s \tilde f(\cdot-B_s) {\rm d}B_s^i$$
is the $\oplus^d L^2(\RR^{d+1})$-valued stochastic integral 
of  $\tilde f=(f/\sqrt\omega\check{)}$.
From this formula we have 
$e^{-tH_{PF}}\Psi(x)=
\Ebb^x[e^{-\int_0^tV(B_s) {\rm d}s} 
J_0^\ast e^{-i\AA({\rm K})}J_t\Phi(B_t)
]$. 
Furthermore
let 
$$K_{PF}=\half (p-A(f))^2$$ be the kinetic term of the Pauli-Fierz model $H_{PF}$. 
It also established that 
$K_{PF}$ is essentially self-adjoint on 
$D(p^2)\cap C^\infty(N)$ when Assumption \ref{h1} is assumed. 
Then 
it follows that 
\eq{hh1}
(\Phi, e^{-tK_{PF}}\Psi)=
\ix \left[e^{-\int_0^tV(B_s) {\rm d}s} 
(\Phi(B_0), e^{-i\A({\rm K})}\Psi(B_t))_{L^2(Q)}
\right], 
\en
where 
$$K=\oplus_{i=1}^ d
\int _0^t  \tilde f(\cdot-B_s) {\rm d}B_s^i$$
is the $\oplus^d L^2(\BR)$-valued stochastic integral.

\subsection{Extended Pauli-Fierz model}
The extended Pauli-Fierz model is defined by 
\eq{hh2}
\widehat H_{PF}=H_{PF}\otimes\one+\one\otimes\hf
\en
as an operator in $\widehat \hhh=\hhh\otimes\fff$. Note that $\fff\otimes\fff\cong \fff(W\oplus W)$. 
Under the identification 
$\widehat \hhh\cong \LR\otimes \fff(W\oplus W)$, 
then 
$$\widehat \HHH_{PF}=\half(p-A_1)^2+\widehat{\hf},$$
where $A_1=A(f\oplus 0)$ and $\widehat {\hf}$ be the second quantization of 
$
\widehat \omega=
\omega \oplus 
\omega$. 
Then the functional integral representation 
of $e^{-t\widehat H_{PF}}$ is a slight modification of that of $e^{-tH_{PF}}$.

Let $\A(F)$ be a 
Gaussian random variables on a probability space $(\tilde Q,\tilde \Sigma,\tilde\mu)$, which  is indexed by 
$F\in 
(\oplus^d L^2(\BR))
\oplus 
(\oplus^d L^2(\BR))
$. 
Let $\A_1(F)=\A(F\oplus 0)$ and $\A_2(G)=\A(0\oplus G)$.  
The  mean of $\A_\#(F)$ is 
zero and the covariance is given by 
\eq{hh11}
\Ebb[\A_i(F)\A_j(G)]= \half \delta_{ij} 
\sum_{\mu,\nu=1}^d (\hat F_\mu, 
\delta_{\mu\nu}^\perp\hat G_\nu),\quad i,j=1,2.
\en
Similar to the Pauli-Fierz model 
we introduce the Euclidean version of $\AA_j$, $j=1,2$.
Let $\widehat J_t=J_t\otimes J_t$ and $\widehat \j_t=\j_t\otimes \j_t$.
Then 
$\widehat J_t: L^2(Q)\otimes L^2(Q)\to 
L^2(Q_E)\otimes L^2(Q_E)$
and $\widehat \j_t:L^2(\BR)\otimes L^2(\BR)\to L^2(\RR^{d+1})\otimes L^2(\RR^{d+1})$
 satisfy 
that 
$\widehat J_t^\ast \widehat J_s=
e^{-|t-s|\widehat {\hf}}$ and 
$\widehat \j_t^\ast \widehat \j_s=
e^{-|t-s|
\widehat\omega(-i\nabla)}$. 
Hence we can see that 
\eq{hh3}
(\Phi, e^{-t\widehat H_{PF}}\Psi)=
\ix \left[e^{-\int_0^tV(B_s) {\rm d}s} 
(\widehat J_0\Phi(B_0), e^{-i\AA_1
({\rm K}_E)}
\widehat J_t\Phi(B_t))
_{L^2(Q_E)\otimes L^2(Q_E)}
\right].
\en

\subsection{Generalization  of extended Pauli-Fierz model}
Let $f$ and $g$ be two cutoff functions and we define 
$\widehat H_{PF}$ 
by 
\eq{hh4}
\widehat H_{PF}=
\half(p-A(f)\otimes\one_\fff -\one_\hhh \otimes A(g))^2+\hf\otimes\one+\one\otimes\hf. 
\en
Hence 
we can see that 
\begin{align}
& (\Phi, e^{-t\widehat H_{PF}}\Psi) \non\\
&\quad =
\ix \left[e^{-\int_0^tV(B_s) {\rm d}s} 
(\widehat J_0\Phi(B_0), 
e^{-i\AA_1({\rm K}_f)-i\AA_2({\rm K}_g)}
\widehat J_t\Phi(B_t))_{L^2(Q_E)\otimes L^2(Q_E)}
\right], \label{hh6}
\end{align}
where 
${\rm K}_h=\oplus_{i=1}^{d}\int_0^t \j_s 
\tilde h(\cdot-B_s) {\rm d}B_s^i$ for $h=f,g$. 
Furthermore 
let 
$$\widehat K_{PF}=
\half(p-A(f)\otimes\one_\fff -\one_\hhh \otimes A(g))^2$$
be the kinetic term of $\widehat H_{PF}$. 
It can be shown that $\widehat K_{PF}$ is essentially self-adjoint on $D(p^2)\cap C^\infty (\widehat N)$. 
We then also have 
\eq{hh7}
(\Phi, e^{-t\widehat K_{PF}}\Phi)=
\ix \left[e^{-\int_0^tV(B_s) {\rm d}s} 
(\Phi(B_0), e^{-i\A_1({\rm K}_F)-i\A_2({\rm K}_G)}
\Phi(B_t))_{L^2(Q)\otimes L^2(Q)}\right].\en

\noindent {\bf Acknowledgments:} 
FH   thanks the hospitality of universit\'e de Paris XI, universit\'e d'Aix-Marseille-Luminy and  universit\'e de
Rennes 1, where part of this work has been  done.

\end{document}